\documentclass[twocolumn,showpacs, showkeys,superscriptaddress,
preprintnumbers,amsmath,amssymb]{revtex4}
\usepackage{graphicx}
\usepackage{dcolumn} 
\usepackage{bm}

\begin{document}
\title{First measurement of the Head-Tail directional nuclear recoil signature at energies relevant to WIMP dark matter searches}% 

\author{S. Burgos}
\affiliation{Department of Physics, Occidental College, Los Angeles, CA  90041, USA.}
\author{E. Daw} 
\affiliation{Department of Physics and Astronomy, University of Sheffield,  Sheffield, S3 7RH, UK.}
\author{J. Forbes}
\affiliation{Department of Physics, Occidental College, Los Angeles, CA  90041, USA.}
\author{C. Ghag} 
\affiliation{School of Physics and Astronomy, University of Edinburgh, Edinburgh, EH9 3JZ, UK.}
\author{M. Gold} 
\affiliation{Department of Physics and Astronomy, University of New Mexico, NM 87131, USA.}
\author{C. Hagemann} 
\affiliation{Department of Physics and Astronomy, University of New Mexico, NM 87131, USA.}
\author{V.A. Kudryavtsev} 
\affiliation{Department of Physics and Astronomy, University of Sheffield,  Sheffield, S3 7RH, UK.}
\author{T.B. Lawson}
\affiliation{Department of Physics and Astronomy, University of Sheffield,  Sheffield, S3 7RH, UK.}
\author{D. Loomba} 
\affiliation{Department of Physics and Astronomy, University of New Mexico, NM 87131, USA.}
\author{P. Majewski} 
\affiliation{Department of Physics and Astronomy, University of Sheffield,  Sheffield, S3 7RH, UK.}
\author{D. Muna}
\affiliation{Department of Physics and Astronomy, University of Sheffield,  Sheffield, S3 7RH, UK.}
\author{A.StJ. Murphy}
\affiliation{School of Physics and Astronomy, University of Edinburgh, Edinburgh, EH9 3JZ, UK.}
\author{G.G. Nicklin}
\affiliation{Department of Physics and Astronomy, University of Sheffield,  Sheffield, S3 7RH, UK.}
\author{S.M. Paling} 
\affiliation{Department of Physics and Astronomy, University of Sheffield,  Sheffield, S3 7RH, UK.}
\author{A. Petkov}
\affiliation{Department of Physics, Occidental College, Los Angeles, CA  90041, USA.}
\author{S.J.S. Plank} 
\affiliation{School of Physics and Astronomy, University of Edinburgh, Edinburgh, EH9 3JZ, UK.}
\author{M. Robinson} 
\affiliation{Department of Physics and Astronomy, University of Sheffield,  Sheffield, S3 7RH, UK.}
\author{N. Sanghi} 
\affiliation{Department of Physics and Astronomy, University of New Mexico, NM 87131, USA.}
\author{D.P. Snowden-Ifft}  
\affiliation{Department of Physics, Occidental College, Los Angeles, CA  90041, USA.}
\email{ifft@oxy.edu}
\author{N.J.C. Spooner} 
\affiliation{Department of Physics and Astronomy, University of Sheffield,  Sheffield, S3 7RH, UK.}
\author{J. Turk} 
\affiliation{Department of Physics and Astronomy, University of New Mexico, NM 87131, USA.}
\author{E. Tziaferi}
\affiliation{Department of Physics and Astronomy, University of Sheffield,  Sheffield, S3 7RH, UK.}

\date{\today}

\begin{abstract}
 We present first evidence for the so-called Head-Tail     
asymmetry signature of neutron-induced nuclear recoil tracks at            
energies down to 1.5 keV/amu using the 1m$^{3}$  DRIFT-IIc dark matter detector.        
This regime is appropriate for recoils induced by Weakly               
Interacting Massive Particle (WIMPs) but one where the differential          
ionization is poorly understood.  We show that the distribution of        
recoil energies and directions induced here by $^{252}$Cf neutrons matches well       
that expected from massive WIMPs.  The results open a powerful new means of     
searching for a galactic signature from WIMPs.   

\end{abstract}

\pacs{Valid PACS appear here}

\keywords{Dark Matter, WIMPs, neutralino, TPC, MWPC, Boulby}

\maketitle

	Determining the nature of dark matter remains an outstanding goal in physics with Weakly Interacting Massive Particles (WIMPs) a leading candidate~\cite{WIMPReview}.   A world-wide effort to observe WIMP-induced nuclear recoils expected in terrestrial materials is underway.  Recently, new results from DAMA, observing all events in a 250 kg NaI array, improve the significance of the signal they claim indicates the presence WIMPs through the small annual modulation of event rate expected due to variation in the WIMP velocity distribution as the Earth orbits the
Sun~\cite{dama2008}.  Annual modulation provides one route towards a galactic signature based on the non-terrestrial nature of WIMPs, allowing some model-independency from particle physics and cosmology assumptions.  However,  it is known that diurnal directional signatures, that provide a definitive detection impossible to mimic by terrestrial backgrounds, could come from a device capable of tracking the direction of WIMP-induced recoil ions, down to low ($\sim$1 keV/amu) energies, event by event~\cite{green}.   

The only method established for observing these powerful but difficult to extract signatures, is to use a low density, large volume, target with sufficient spatial resolution to do the necessary tracking~\cite{PRD}.  The DRIFT (Directional Recoil Identification From Tracks) experiment at Boulby mine~\cite{DI, DIIaneut},  is designed to achieve this using a low pressure CS$_{2}$ gas Negative Ion Time Projection Chamber (NITPC).    A range component signature has recently been demonstrated with the DRIFT-IIc detector utilizing  $\sim$1.5 keV/amu S recoils~\cite{signaturesI}.  Previous studies have shown that only $\sim$100 WIMP events are, in principle, needed to demonstrate the galactic nature of such a range component signature~\cite{green}.  However, reconstruction of the full recoil direction vector, to distinguish the ``tail" from the ``head" (we define the tail as the point closest to the initial collision), as opposed to the direction axis alone, critically influences this sensitivity,  in principle lowering the required number of events to $\sim$10. 

	Confirmation of the head-tail effect is of major interest. While the means to achieve this is through measurement of the variation in ionization density along the track, no previous experimental studies have been made in relevant energy range ($\sim$ 1~keV/amu) and theoretical predictions are known to be unreliable. However, success would open prospects for a new generation of experiment with sensitivity to the galactic WIMP velocity distribution.
	
	Based on tests  with a 1\,m$^3$ DRIFT module with realistic energy threshold  and size for WIMP searches, we present here first positive results in both these areas.  This is achieved by measuring the integrated ionization loss split between the head and tail of S recoil tracks down to 1.5 keV/amu, induced by $^{252}$Cf  neutrons.  These measurements show that reconstruction of the $^{252}$Cf neutron recoil ion direction vector is possible.  We show also that the recoil directions resulting from exposure to the $^{252}$Cf source models well that expected from a Maxwellian dark-matter halo and hence that it is possible to measure the variation of interaction rate with WIMP recoil direction due to rotation of the Earth.   These are significant milestones toward realizing a large-scale directional detector, bringing closer the prospect of a definitive identification of dark matter, including study of the halo WIMP velocity distribution and independent investigation of DAMA.   

\begin{figure}
\includegraphics[width=0.45\textwidth]{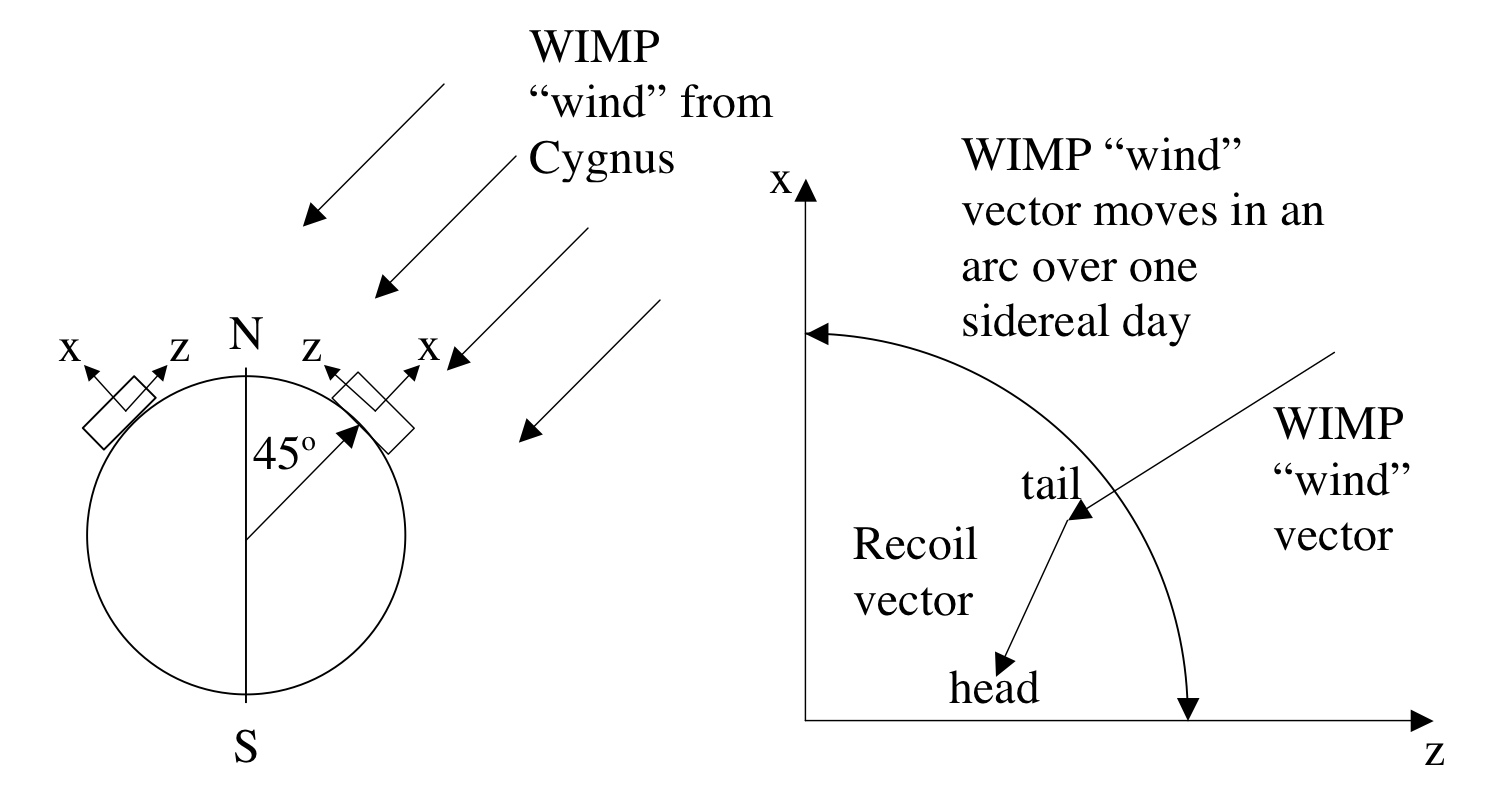}
\caption{\label{fig:wimpwind} (a) WIMP ``wind"" from constellation Cygnus.  (b) A detector at latitude 45$^\circ$ sees the ``wind'' vector oscillate from horizontal (South) to vertical over a sidereal day.}
\vspace{-5mm}
\end{figure}

	The basis of the signatures (see Figure \ref{fig:wimpwind}) is the Earth's motion through a stationary Galactic WIMP halo to produce a WIMP ``wind" from the direction Cygnus, declination $\sim$45$^\circ$.  Recoil tracks produced will preferentially be aligned with this wind vector~\cite{PRD}.    A directional detector on Earth at a latitude ~$\sim$45$^\circ$, like Boulby, will thus see the WIMP wind vector oscillate over a sidereal day, from pointing south to pointing to the Earth's center, repeating each sidereal day and going rapidly out of phase with the terrestrial day.   	
	
	The recoils produced are extremely low energy and their differential ionization as they slow down is unclear.  Electronic stopping power, energy loss to electrons, is expected to decrease from tail to head~\cite{Hitachi}, suggesting more ionization at the tail than head. However, as the ion slows, nuclear stopping, energy loss to nuclear recoils, also becomes important.    Also, ions at the end of their range scatter frequently and so tend to Òball upÓ, losing their last energy over a small region, possibly creating an ionization spike at the track-head, observable depending on the topology relative to the readout.  These issues are explored theoretically for DRIFT in~\cite{pawel}.  Even if these effects yield a theoretical head-tail asymmetry, the detector resolution and diffusion to a readout plane could swamp the effect and it anyway may be too small to be of use in a full sized detector.  The DRIFT-II dark matter detectors, as operating low pressure gaseous devices~\cite{DIIaneut},  provide a unique testbed for simultaneously determining whether the head-tail effect exists and whether this is of use in operational full-sized experiments.

	The design and operation of DRIFT-II has been presented elsewhere ~\cite{DII,Lawson, DIIaneut}. The specific module used here, DRIFT-IIc, is essentially identical to IIa,b used for underground dark matter runs.   Briefly, the device comprises a 1.5 m$^3$ low background stainless steel vessel filled with 40 Torr CS$_{2}$ and containing two identical back-to-back TPCs with a common 1 m$^{2}$ central cathode plane. Recoil events can form in two identical drift regions of 1 m$^{2}$ by 50 cm depth.  Ionization tracks are drifted in each TPC away from the central cathode, to be read out by either of two MWPCs. We defined here the direction perpendicular to the MWPCs as the z-direction, this being horizontal, the +z direction defined as left to right viewed from the front with the origin on the central cathode.  Each MWPC comprises 448 grid (y-direction) and 448 anode wires (x-direction) of 2 mm pitch, grouped into 8 signal lines per readout plane.  Eight adjacent lines (either anode or grid) therefore sample a distance of 16 mm in x and y.  52 wires at the edges provide veto signals against side events, such as alphas~\cite{Lawson}, determining a final fiducial volume of 0.80 m$^{3}$ (134 g of CS$_{2}$).  Each drift volume is instrumented with a retractable $\sim$100 $\mu$Ci $^{55}$Fe source for gain and functionality monitoring.

\begin{figure}
\includegraphics[width=0.45\textwidth]{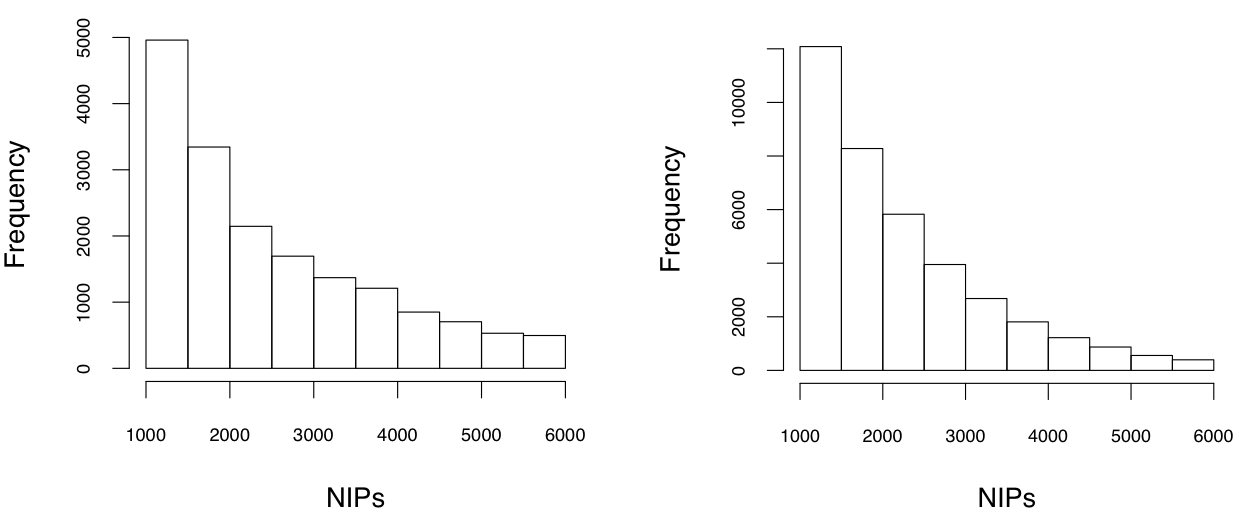}
\caption{\label{fig:nipsspectrum}  Predicted NIPs spectrum for (left) neutron induced S recoils; and (right) from 1000 GeV WIMPs  (using GEANT).}
\vspace{-3mm}
\end{figure}

\begin{figure}
\includegraphics[width=0.45\textwidth]{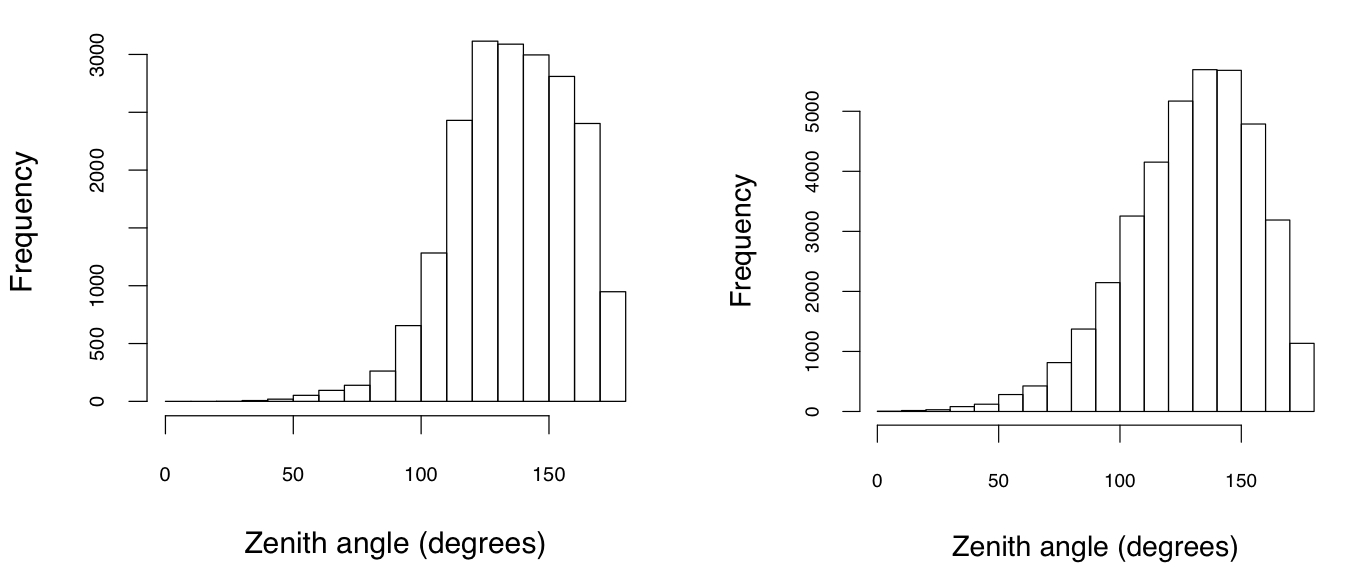}
\caption{\label{fig:mcresults} (left) GEANT produced spectrum of $>$ 1000 NIPs S recoil zenith angles (z axis) from -z directed $^{252}$Cf neutrons; and (right) CYGNUS Monte Carlo equivalent for -z directed 1000 GeV WIMPs.  The WIMP induced recoils are peaked slightly higher (closer to neutron direction).}
\vspace{-3mm}
\end{figure}

	The basis of the head-tail test here was use of a 202\,$\mu$Ci $^{252}$Cf source placed in turn to produce neutrons headed predominantly in the +x, -y, +z and -z directions, positioned always at 330 cm from the detector's center so that the angular spread of the interacting neutrons (excluding scattering) was $\sim$10$^{\circ}$.  Simulations of neutron and WIMP induced recoils using a GEANT detector simulation and our NeuRec/Cygnus Monte Carlos~\cite{neutron} have shown such exposures mimic WIMPs quite well.   Certainly the average energy  for C recoils from $^{252}$Cf neutrons is larger than expected from WIMPs.  However, as shown in~\cite{signaturesI} because of their lower stopping power C recoils can be rejected in favor of S by typically ~200:1. Thus, the results here apply to S only which is fortunate since S recoils are more important for spin-independent WIMP searches due to the A$^{2}$ cross section dependence.  For S, the comparison between neutron and massive WIMP induced events is quite favorable.  Figure \ref{fig:nipsspectrum} compares the S recoil spectrum from $^{252}$Cf neutrons (using GEANT) and 1000 GeV WIMPs (using the CYGNUS in-house Monte Carlo).  Figure \ref{fig:mcresults} shows comparison of the expected S recoil direction vectors from neutrons along the positive z axis (-z neutrons moving predominantly right to left) with those from WIMPs when Cygnus is on the positive z axis.  These plots reveal that the neutrons produce only a slightly harder spectrum, with the WIMP induced recoils slightly more forward peaked.  The S recoils produced and analyzed in this experiment were a reasonable representation of S recoils from massive WIMPs.

	Based on this, the head-tail signature in either +z or -z neutron exposure, should, if present, be seen through production of a different degree of ionization at the head or tail revealed as an asymmetry in the time distribution of ionization arrival at the MWPC planes.  For instance, if the effect produces more ionization at the tail than the head then for a -z directed neutron run, events on the right TPC will have, on average, more ionization earlier than later.  Events on the left  TPC will have the reverse.   For the +z run the effect will be opposite.  For x and y runs, with neutrons parallel to the MWPCs, the ionization on average will appear the same at the beginning and end.  Note how with back-to-back TPCs identical but for opposing fields, DRIFT is well designed to control systematics in head-tail studies.

	To quantify the asymmetry a simple event-by-event analysis has been developed where the voltage integral over the first and second half of each accepted ionization time profile is calculated and the ratio taken as an asymmetry measure.  In DRIFT, the voltage integral is proportional to the Number of Ion Pairs (NIPs), convertible to event energy using source calibrations.  For instance, 1000 NIP S recoils have energies of 47 keV~\cite{neutron}.  The asymmetry quantity used is thus: {\it NIPsRatio} =  {\it NIPs$_{1}$}/{\it NIPs$_{2}$}, where indices refer to the first and second half of the pulse.   Determination of {\it NIPsRatio} follows by locating the highest voltage and using this to find the first time that the waveform crosses 25\% (chosen for optimum signal-to-noise) of this value on either side.  A study of various values instead of 25\% revealed it to have the highest signal to noise ratio for this study.  This process minimizes potential biases from waveform noise and baseline shifts. The middle of the track is defined as midway between these times.  NIPs$_{1}$ is the integral voltage above 25\% of the peak value at minimum time to the middle and NIPs$_{2}$ the integral from there to 25\% of peak value at maximum time.  

	Prior to this a first stage analysis is used to extract the recoil signals from unwanted backgrounds, like sparks and alphas. For details of cuts, efficiencies and track reconstruction used see~\cite{signaturesI} and for general concepts~\cite{neutron}. Particular issues here were, for instance, the software threshold, chosen as 50 ADC counts to include ionization on neighboring anode wires.   For large events, induced pulses  were found to trigger the analysis, a new cut removed these.   A set of 22 parameters are calculated for each output line to generate noise cuts and properly calculate event ionization.  For instance, accepted events are required to fall within defined time windows on anode and grid; to deposit the same energy on grid and anode to 10\%; not to be clipped by the digitizers; to be within the fiducial volume; to appear in only one TPC; and to trigger adjacent lines but not all lines.  Sparks and MWPC events are removed by a FWHM cut on individual lines.  

\begin{figure}
\includegraphics[width=0.45\textwidth]{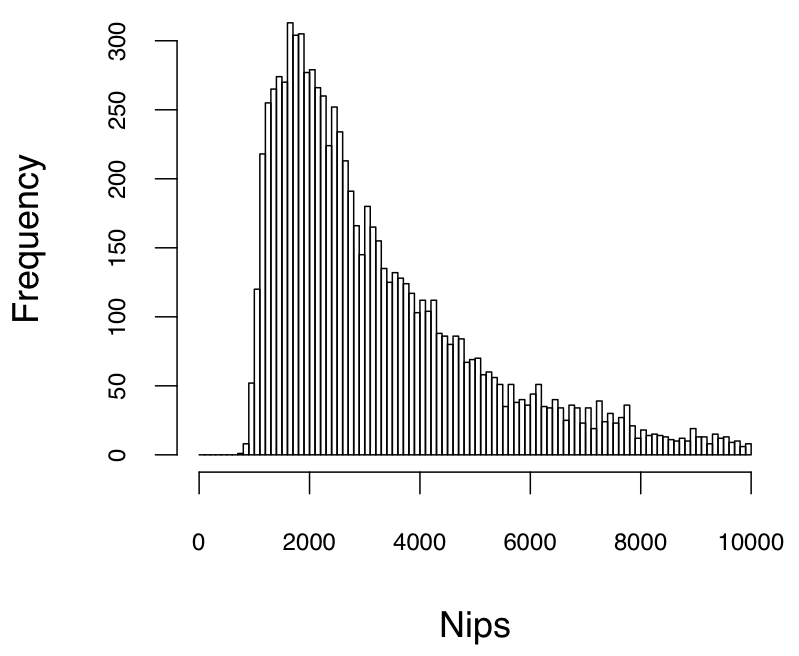}
\caption{
\label{fig:espect} Example neutron S recoils NIPs spectrum. 
\vspace{-3mm}}
\end{figure}

\begin{figure}[h!]
\begin{center}
\includegraphics[width=0.45\textwidth]{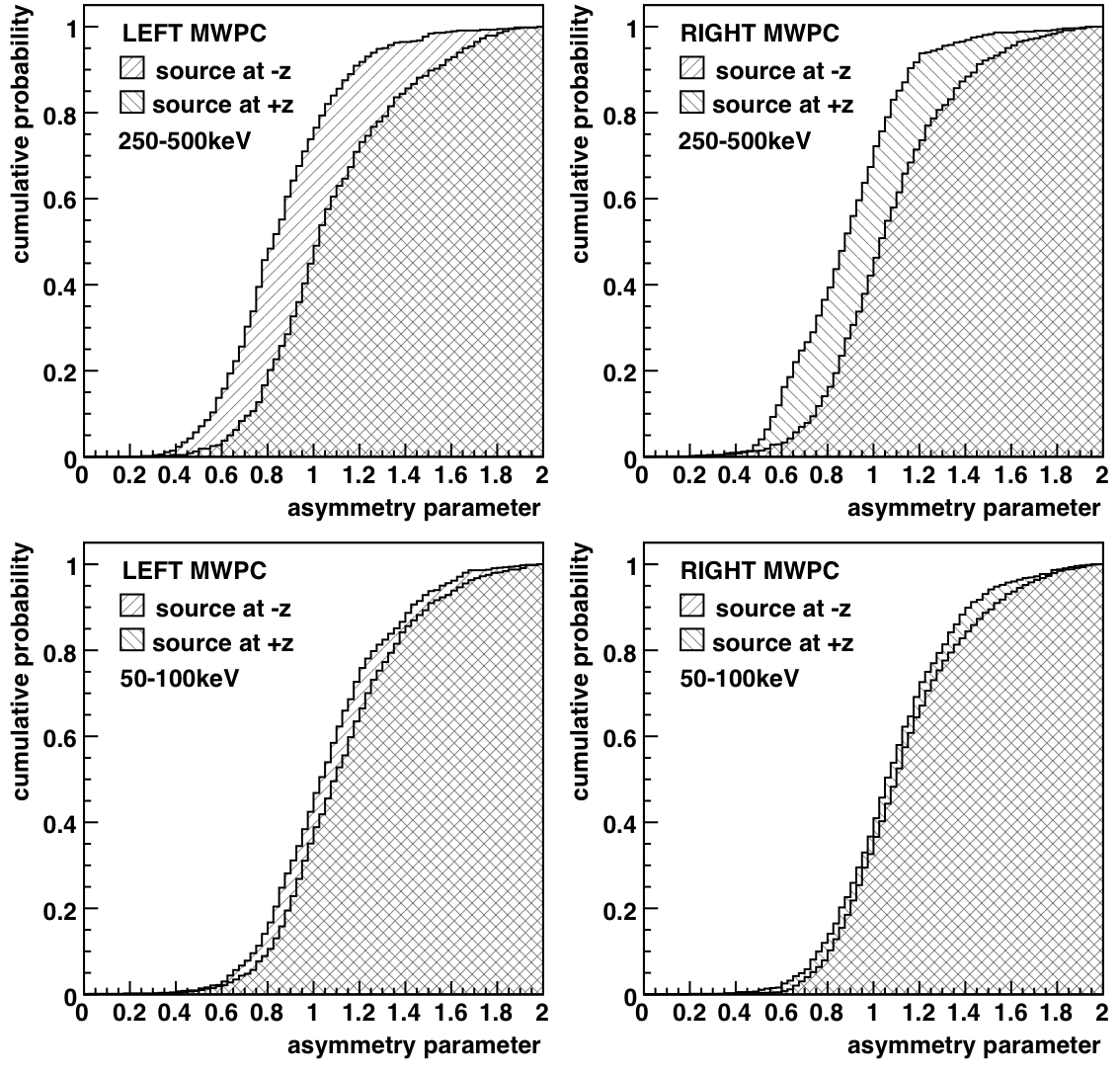}
\caption{\label{fig:inthist} Cumulative asymmetry
parameter ({\it NipsRatio}) for $\pm\mathrm{z}$ direction on each TPC
for 250-500\,keV and 50-100\,keV recoils. 
\vspace{-3mm}}
\end{center}
\end{figure}

\begin{figure}[h!]
\includegraphics[width=0.45\textwidth]{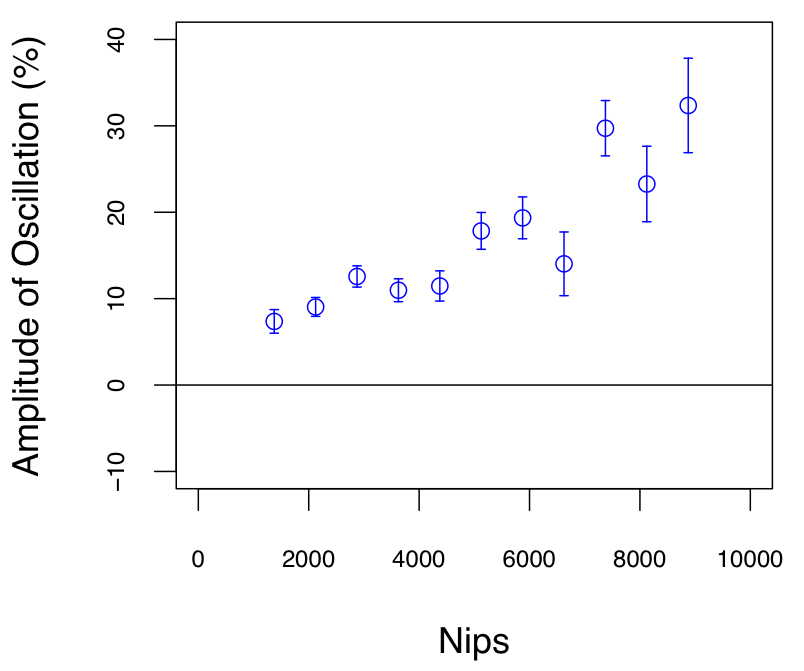}
\caption{\label{fig:asymstats} Head-Tail signal - oscillation amplitude vs. NIPs.}
\vspace{-3mm}
\end{figure}

	From this we show in Figure \ref{fig:espect} a typical NIPs energy spectrum for neutron-induced S recoils.  Here, the hardware threshold and the analysis cuts produced an effective threshold of $\sim$1000 NIPs, 47 keV S recoil ~\cite{neutron}.    Table \ref{tab:runstats} and Figures \ref{fig:inthist} and \ref{fig:asymstats} then summarize key results from +x, -y, +z and -z directed neutron runs.  In Table 1,  column 2 shows the number of events passing cuts for each exposure with direction as in column 1.   The means of the {\it NIPsRatio} distribution for left and right TPC are in columns 3, 4, their difference and the significance (difference divided by combined error) in columns 5, 6.

\begin{table*}
\caption{\label{tab:runstats} Directional run statistics}
\begin{ruledtabular}
\begin{tabular}{cccccc}
 Run&N&Left&Right
&Left-Right&S\\ \hline
+x & 8673 & $1.074\pm{0.008}$&$1.069\pm{0.004}$&$0.005\pm{0.009}$&0.549\\
-y & 5859 & $1.082\pm{0.006}$&$1.083\pm{0.006}$&$-0.001\pm{0.009}$&-0.121\\
 +z & 5829 & $1.145\pm{0.009}$&$1.007\pm{0.006}$&$0.14\pm{0.01}$&13.4\\
-z & 8755 & $0.995\pm{0.006}$&$1.143\pm{0.005}$&$-0.147\pm{0.008}$&-19.2\\
 - & - & - & - & - &-\\
 Optimal (+z and -z) & 14584 & -&-&$0.143\pm{0.006}$&23.8\\
 Anti-optimal (+x and -y) & 14532 & -&-&$0.005\pm{0.006}$&0.756\\
\end{tabular}
\end{ruledtabular}
\vspace{-2mm}
\end{table*}
		
		Consider the +x run (Table \ref{tab:runstats}). It is expected, even in the presence of a head-tail effect, that the average {\it NIPsRatio}$=$1.0 since equal numbers of events are headed towards or away from the detectors.  In fact the mean ratio is not 1.0, within errors.  This systematic shift is due to amplifier overshoot that tends to shorten the second half of the event.  The left and right mean {\it NIPsRatios} are, however, in statistical agreement (columns 5, 6), suggesting that the electronic shaping on both sides is identical.  As confirmation the +y run shows the same result within the errors.   Now comparing the +z run with the +x and -y runs we can see that events on the left side have a larger {\it NIPsRatio}, events on the right, smaller.  The mean difference is 14\% with significance 13.4.  The sign here, left {\it NIPsRatio} bigger than right, suggests more ionization at the tail than head in agreement with the theory presented in~\cite{pawel}.  As an important check it can be seen from the -z run that the results for the left agree with those for the right from the +z run, as expected.  The sign of the difference (column 5), reverses but is otherwise statistically equivalent to the -z run.   Figure \ref{fig:inthist}  illustrates the effect further.   Here, integration histograms of the asymmetry parameter for 250-500\,keV and 50-100\,keV events for each MWPC are shown, normalized to the total event count. Plots with the source in the +z and -z positions are overlaid. The difference between +z and -z is clearly visible, growing smaller at lower energy.
		
		To provide the best measurement of this effect events on the left(right) of the +z(-z) runs for which the {\it NIPsRatio} measures the ratio of the ionization at the head of the track to the tail, were combined, as were the events on the right(left) of the +z(-z) runs for which the {\it NIPsRatio} measures the ratio of the ionization at the head to the tail.  The difference in these average ratios, labeled Optimal in Table 1, has a high significance of 23.8. As a final check the same procedure was applied to the ÒAnti-OptimalÓ +x and -y runs, yielding a null result (columns 5, 6).   To aid understanding the oscillation amplitude vs. NIPs was calculated as for the Optimal runs.  As clearly seen in Figure \ref{fig:asymstats} the magnitude  increases with increasing energy and, by extrapolation, looks likely to remain significant even below 1000 NIPs. 
		
		In conclusion, it has been demonstrated using a m$^{3}$-scale DRIFT directional dark matter detector that neutron-induced S recoils, with properties close to those from massive WIMPs, display a clear head-tail asymmetry at low energy.  Even analysis using non-optimized electronics and simple parameterization comparing ionization deposited in the first and second half of the recoil tracks, indicates that head-tail asymmetry can be observed equivalent to a modulation, as the WIMP directions change over a sidereal day, of 14\% at current threshold.  In~\cite{signaturesI} we show how the orientation of recoil tracks, parameterized by the ratio $\Delta$z/$\Delta$x, will also oscillate over a sidereal day.   The prospect is now open of combining these signals, leading to a new and more powerful route to identifying or studying the galactic nature of WIMPs.  Analysis of this will be the subject of future DRIFT work. 
	
	We thank J. Martoff for useful discussions, CPL for access to Boulby and NSF and ILIAS for financial support.

\end{document}